# Uniaxial extensional viscosity of semidilute DNA solutions

Sharadwata Pan[1,2,3], Duc At Nguyen[3], P. Sunthar[1,2], T. Sridhar[1,3] and J. Ravi Prakash[1,3,*]

[1]IITB-Monash Research Academy, Indian Institute of Technology Bombay, Powai, Mumbai 400076, India

[2]Department of Chemical Engineering, Indian Institute of Technology Bombay, Powai, Mumbai 400076, India

[3]Department of Chemical Engineering, Monash University, Melbourne, VIC 3800, Australia

## Abstract

The extensional rheology of polymer melts and dilute polymer solutions has been extensively examined through experiments and theoretical predictions. However, a systematic study of the extensional rheology of polymer solutions in the semidilute regime, in terms of examining the effects of concentration and molecular weight, has not been carried out so far. Previous experimental studies of the shear rheology of semidilute polymer solutions have demonstrated that their behaviour is distinctively different from that observed in the dilute and concentrated regimes. This difference in behaviour is anticipated to be even more pronounced in extensional flows, which play a critical role in a number of industrial contexts such as fiber spinning and ink-jet printing. In this work, the extensional rheology of linear, double-stranded DNA molecules, spanning an order of magnitude of molecular weights (25–289 kilobasepairs) and concentrations (0.03–0.3 mg/ml), has been investigated. DNA solutions are now used routinely as model polymeric systems due to their near-perfect monodispersity. Measurements have been carried out with a filament stretching rheometer since it is the most reliable method for obtaining an estimate of the elongational stress growth of a polymer solution. Transient and steady-state uniaxial extensional viscosities of DNA dissolved in a solvent under excess salt conditions, with a high concentration of sucrose in order to achieve a sufficiently high solvent viscosity, have been determined in the semidilute regime at room temperature. The dependence of the steady state uniaxial extensional viscosity on molecular weight, concentration and extension rate is measured with a view to determining if data collapse can be observed with an appropriate choice of variables. Steady state shear viscosity measurements suggest that sucrose-DNA interactions might play a role in determining the observed rheological behaviour of semidilute DNA solutions with sucrose as a component in the solvent.

----------------------------------------------------------------------

*Corresponding author, E-mail: ravi.jagadeeshan@monash.edu

# 1. Introduction

Several significant industrial processes such as fiber spinning, ink-jet printing, the extrusion of polymeric materials, and applications such as coatings, turbulent drag reduction and lubrication, predominantly involve the extensional mode of deformation (McKinley and Sridhar, 2002). Unlike shear flows, however, in which vorticity leads to the tumbling of polymer molecules and consequently, an incomplete extension of polymer chains, extensional flows are irrotational and capable of completely unravelling and orienting flexible chains (McKinley and Sridhar, 2002). As a result, the elongational viscosity of a polymeric liquid is a crucial material function, which significantly impacts far from equilibrium behavior. The rheological behavior of dilute and concentrated polymer solutions and melts in extensional flows, both at steady state and in transient elongational flow, has been extensively investigated through experiments, simulations and theoretical predictions [Tirtaatmadja and Sridhar, 1993; Gupta et al., 2000; McKinley and Sridhar, 2002; Bhattacharjee et al., 2003; Sunthar et al., 2005; McKinley and Hassager, 1999; Bach et al., 2003; Wang et al., 2011; Wang and Wang, 2011; see also the recent reviews by Shaqfeh, 2005; Larson and Desai, 2015; Schroeder, 2018; and Prakash, 2019]. Progress has been possible in both these extreme regimes of concentration since it is sufficient to focus on single chain dynamics in order to obtain an understanding of the rheological behaviour. On the other hand, the semidilute regime of polymer solutions involves many-body interactions, that lead to complex and rich behavior. A thorough experimental study aimed at understanding the bulk extensional rheology of polymer solutions in the semidilute regime is currently lacking, and a systematic examination of the effects of concentration and molecular weight is clearly required for a proper rheological characterization of these systems.

Over the years, there have been attempts to characterize the bulk extensional rheology of semidilute polymer solutions based on various devices that exploit the capillary-driven thinning of the neck of an asymmetric liquid bridge, including capillary breakup (Clasen, 2010), jet breakup (Christanti and Walker, 2001), and lab-on-a-chip microfluidic prototypes such as dripping-onto-substrate rheometry (Dinic et al., 2017a, Dinic et al., 2017b, Dinic and Sharma, 2019). In these measurements, the extensional viscosity is obtained indirectly from a balance between the elastic and capillary stresses in the elastocapillary regime (Dinic et al., 2017b). On the other hand, the Filament Stretching Rheometer (FSR) enables a direct measurement of the extensional viscosity by measuring both the stress in a filament undergoing uniaxial extensional flow, and the local strain rate. Filament stretching rheometry has been thoroughly reviewed by McKinley and Sridhar (2002). The current FSR technique creates a perfect, standardized, uniaxial extensional flow and delivers the most consistent quantification of the extensional stress growth in a polymer solution (Anna et al., 2001; McKinley and Sridhar, 2002; Sunthar et al., 2005). In our group, filament stretching rheometry has been used previously to characterize dilute, monodisperse, polystyrene solutions at various concentrations and molecular weights (Gupta et al., 2000), the dynamics of entangled polymer solutions under both planar as well as uniaxial extensional deformations (Sridhar et al., 2013; Nguyen et al., 2015), and dilute DNA solutions under uniaxial extensional flow (Sunthar et al., 2005). In the current work, a custom-built filament stretching rheometer is employed to characterize the uniaxial extensional stress growths of semidilute DNA solutions.

Due to their near ideal monodispersity, propensity to get stained with considerable ease, and behavior that has been shown to be identical to charge neutral synthetic polymers under high salt conditions (Pan et al., 2014a,b), the suitability of DNA solutions as model solutions to investigate important questions in polymer solution physics in general, and oscillatory and steady shear rheology in particular, have been well documented in the dilute, semidilute and entangled regimes, for both single and double-stranded configurations (Pecora, 1991; Brockman et al., 2011; Pan et al., 2014a,b; Regan et al., 2016; Pan et al., 2018, Goudoulas et al., 2018). Hsiao et al. (2017a) have recently directly observed the dynamics of single $\lambda$-DNA chains, in dilute and semidilute unentangled solutions, under planar extensional flow. Their study showed that the concentration of DNA significantly influences its stretching dynamics in the semidilute regime, which are qualitatively distinct compared to that in the dilute regime. In this work, we have investigated the extensional rheology of linear DNA molecules in a wide range of molecular weights (25 to 289 kbp) and concentrations. The current study is the first attempt to quantify the transient and steady state uniaxial elongational stress growth in semidilute, unentangled DNA solutions, as a function of chain length and concentration.

Recent experimental work in our group on the shear flow of semidilute, unentangled DNA solutions has revealed that a clear understanding of the concentration and temperature dependence of nonlinear viscoelastic properties can be achieved in terms of the Weissenberg number Wi, the scaled concentration $c/c*$, and the solvent quality parameter $z$ (Pan et al., 2014b; Pan et al., 2018). The objectives of the current work are two-fold: first, to generate a set of benchmark data for semidilute DNA solutions across a range of molecular weights and concentrations in elongational flow and second, to examine if the concentration dependence of the extensional viscosity in the semidilute regime can be interpreted in terms of appropriate variables that lead to data collapse, which would be helpful for the development of predictive models. The remaining part of the paper is organized as follows. In Sec. 2, we discuss sample procurement and preparation, and the experimental protocol, including the custom-made filament stretching rheometer setup. In Sec. 3, the shear characteristics and the extensional flow properties of semidilute DNA solutions are deliberated. Section 3.1 considers steady shear viscosities, and compares measurements made in solvents with and without sucrose. In section 3.2, the impact of DNA molecular weight and concentration on the steady state uniaxial extensional viscosities is discussed. Finally, the principal conclusions from the current work are summarized in Sec. 4.

## 2. Methodology

### 2.1. Procurement and preparation of samples and solvent composition

Four linear, double-stranded DNA samples were used in the current work, with characteristic properties listed in Table 1. Linear genomic DNA of $\lambda$-phage virus (#N3011L; size 48.5 kilobasepairs or kbp) and T4 phage virus (#314-03973; size 165.6 kbp) were commercially procured from New England Biolabs (U.K.) and Nippon Gene (Japan), respectively. 25 kbp and 289 kbp DNA were originally obtained from Smith's group at UCSD, as cultures of *Escherichia coli* (*E. coli*) in agar stabs containing these as specific double stranded DNA constructs. The details about preparation of the 25 and 289 kbp DNA are mentioned elsewhere (Pan et al., 2014b). After procurement the DNA was extracted, linearized and purified according to the protocol suggested in Laib et al., 2006, adhering to standard molecular biology protocols (Sambrook and Russell, 2001). Briefly, *E. coli* cells containing the 25 and 289 kbp (double stranded, circular) DNAs were grown for 16–18 hours at 37°C with vigorous shaking (200 rpm) in standard Luria Bertani (LB) broth – Miller (#L3022, Sigma-Aldrich) supplemented with 0.0125 mg/ml Chloramphenicol or CAM (#C0378, Sigma-Aldrich) and 0.01% L-arabinose (#A3256, Sigma-Aldrich) for 25 kbp and 0.0125 mg/ml CAM, 0.05 mg/ml Kanamycin (#K1377, Sigma-Aldrich) and 0.01% L-arabinose for 289 kbp. The arabinose acts as an inducer for the extra origin of replication inserted into the 25 and 289 kbp fragments, primarily to overcome the problem of extremely low copy number (~1 or 2 copies per cell) (Laib et al., 2006). This gives a higher yield of these two DNAs than usual. The cells were harvested and cell wall lysed by the alkaline-lysis method. The undesirable contaminants in the form of proteins, RNA and genomic DNA were removed using Phenol (#P4557, Sigma-Aldrich), RNaseA (#R6513, Sigma-Aldrich) etc. and the circular DNAs were precipitated with ethanol (#E7023, Sigma-Aldrich). The purified double stranded DNAs were linearized with ApaI (#R0114L, New England Biolabs) for 25 kbp and Mlu I (#R0198L, New England Biolabs) for 289 kbp which contain unique sites in the 25 and 289 kbp DNA sequences, respectively (Laib et al., 2006). The linearized DNAs were subjected to phenol-chloroform (#1024452500, Merck) extraction and ethanol precipitation and finally dissolved in excess milli-Q grade water.

**Table 1.** Characteristic properties of the four double-stranded DNA molecular weights studied in the current work (reproduced from TABLE I. of Pan et al. (2014b). The contour length is estimated using the expression: $L$ = number of base-pairs (bp) × 0.34 nm; the molecular weight is calculated from: $M$ = number of bp × 662 g/mol (where the base-pair molecular weight has been calculated for a sodium-salt of a typical DNA base-pair segment); the number of Kuhn steps from: $N_k = L / (2P)$ (where $P$ is the persistence length, which is taken as 50 nm, considering excess salt conditions), and the radius of gyration at the $\theta$-temperature is calculated based on: $R_g^\theta = L / 6N_k$.

| DNA size (kilobasepairs or kbp) | Nomenclature | $M$ (× 10⁶ g/mol) | $L$ (μm) | $N_k$ | $R_g^\theta$ (nm) |
|---|---|---|---|---|---|
| 25 | -- | 16.6 | 9 | 85 | 376 |
| 48.5 | $\lambda$-DNA | 32.1 | 16 | 165 | 524 |
| 165.6 | T4-DNA | 110 | 56 | 563 | 969 |
| 289 | -- | 191 | 98 | 983 | 1280 |

The final DNA solutions for 25 kbp and 289 kbp were prepared by adding desired volumes of a solvent containing Tris (#T1503, Sigma-Aldrich), EDTA (#E6758, Sigma-Aldrich), sucrose (#S0389, Sigma-Aldrich) and NaCl (#S5150, Sigma-Aldrich), with the composition specified in the caption to Table 2, and by evaporating out the excess water. This was done to ensure the efficient dissolution of DNA. An identical solvent and procedure for dissolving the DNA pellet after precipitation, and for preparing subsequent dilutions, was used for the $\lambda$-DNA and T4-DNA solutions as well. It is worth noting that solvents with the same reagents but slightly different compositions have been used in our group previously to measure the elongational viscosity of dilute DNA solutions (Sunthar et al., 2005) and by Hsiao et al., (2017a) in their planar extension flow studies of semidilute DNA solutions in a cross-slot device. Since the solvent predominantly contains sucrose, it is considerably more viscous than the solvent used by Pan et al. (2018) to study the shear rheology of dilute and semidilute DNA solutions, where the DNA was dissolved in an aqueous buffered solvent without sucrose. For ease of referring to these two solvents subsequently, we use the following acronyms: WS (for the solvent *With Sucrose*) and NS (for the solvent with *No Sucrose*). The high viscosity of the WS solvent facilitated the extension of the DNA filaments with prolonged relaxation times. The solvent viscosities were measured at different temperatures using a HAAKE MARS rheometer (Thermo Fisher Scientific), and are listed in Table 2.

**Table 2**: Shear viscosity of the solvent ($\eta_s$) used in the current work at various temperatures ($T$). The water-based solvent consisted of 10 mM Tris, 1 mM EDTA, 0.5 M NaCl, and 61.2 wt.% sucrose. For ease of reference, the solvent is referred to as WS (with sucrose).

| $T$ (±0.5 °C) | $\eta_s$ (±0.0005 Pa.s) |
|---|---|
| 15 | 0.085 |
| 20 | 0.065 |
| 21 | 0.061 |
| 30 | 0.0399 |
| 35 | 0.0315 |

**2.2. Concentration and purity of DNA solutions**

For $\lambda$-phage and T4-phage genomic DNAs, company specified values of 0.5 mg/ml and 0.24 mg/ml, respectively, were considered as absolute DNA concentrations. It is expected that these DNA samples possess reasonably high degrees of purity. For the 25 kbp and 289 kbp linear DNAs, their concentrations were determined to be 0.272 mg/ml and 0.012 mg/ml, respectively, from agarose gel electrophoresis by comparing with a standard DNA marker (#N0468L, New England Biolabs). Also, the purity of the 25 kbp and 289 kbp DNA samples were assessed by UV-VIS spectrophotometry (#UV-2450, Shimadzu). The $A_{260}/A_{280}$ ratios were 1.91 and 1.86 for 25 and 289 kbp DNA respectively, which indicates acceptable purity for DNA samples, though it is largely an assumption (Laib et al., 2006). The $A_{260}/A_{230}$ ratios were 2.2 and 2.1 for 25 and 289 kbp DNA respectively, which indicates absence of organic reagents like phenol, chloroform etc. (Sambrook and Russell, 2001).

**2.3. The filament stretching rheometer and extensional rheometry**

A Filament Stretching Rheometer (Tirtaatmadja and Sridhar, 1993; Gupta et al., 2000; Sunthar et al., 2005) has been used for all extensional viscosity measurements. The instrument has a very small sample requirement (minimum 10 μl), which is ideal for measuring DNA solutions. The measuring principle of the FSR has been detailed in earlier studies (Gupta et al., 2000; McKinley and Sridhar, 2002), and standardized by Anna et al. (2001), while the theory of uniaxial extensional rheometry has been discussed by Tirtaatmadja and Sridhar (1993) and McKinley and Sridhar (2002). Briefly, the DNA samples are placed between two plates initially at rest and consequently moved in opposite directions at a controlled exponential rate. This produces an elongated liquid bridge that experiences a uniaxial extensional flow close to its midpoint, with a fixed strain-rate (Sridhar et al., 1991; Tirtaatmadja and Sridhar, 1993). The force needed for the separation depends on the stress caused by linear DNA molecules being extended from their equilibrium coil-like shape to elongated shapes. The stress is acquired by measuring this force at the end plates (Sunthar et al, 2005). By carefully choosing the extension rate for the solvent used in this study, the polymer contributions to the stress from other factors such as gravity, surface tension, and inertia were isolated, as suggested elsewhere (McKinley and Sridhar, 2001). The elongational stress growth coefficients (or the extensional viscosities) of the DNA solutions were obtained at different strain rates by a master-curve technique (Gupta et al., 2000). All experiments were conducted at a constant strain rate based on the mid-point diameter and carried out at room temperature ($21 \pm 0.5$ °C).

## 2.4. Shear rheometry

As part of the study, steady state shear viscosities for all the DNA samples were also measured in the solvent WS at different temperatures (15–30 °C) and concentrations (0.03–0.3 mg/ml) using a Contraves Low Shear LS 30 viscometer [$0.01 <$ shear rate $\dot{\gamma} < 100$ s$^{-1}$, cup-and-bob (1T/1T) geometry]. Details of the rheometer, the measuring principle, temperature sensitivity, the shear rheometry procedure, precautions taken during measurements, instrument calibration, the shear rate range employed, sample equilibration time, the dependence on rheometer geometry etc., have all been reported in detail in our earlier work (Pan et al., 2014a,b; Pan et al., 2018). A continuous shear ramp was avoided during the measurements. Prior to making solutions, $\lambda$-DNA and T4 DNA were kept (at their maximum concentrations) at 65°C for 10 minutes and immediately put in ice for 10 minutes, to prevent aggregation of long DNA chains (Heo and Larson, 2005; Hsiao et al., 2017a). The instrument was calibrated with appropriate Newtonian Standards with known viscosities (from 100-1000 mPa.s at 20°C) before measuring actual DNA solutions. Values obtained fall within 5% of the company specified values. At each shear rate, a delay of 5−15 minutes was employed so that the DNA chains have sufficient time to relax to their equilibrium state.

## 3. Results and Discussion

### 3.1. Steady shear viscosity

Steady shear viscosities for three DNA molecular weights (25, 48.5 and 165.6 kbp), dissolved in solvent WS, were measured at various values of shear rate at different temperatures and concentrations. The experimental data covers a wide range of molecular weights ($1.6 \times 10^7$ to $1.9 \times 10^8$ g/mol) and concentrations (0.03 to 0.3 mg/ml). The shear rate dependence of the measured steady shear viscosity η for λ-DNA and T4 DNA solutions are shown in Fig. 1. The shear viscosities at each concentration in the low shear rate plateau region were least-square fitted with horizontal lines and extrapolated to zero shear rate, as detailed in our earlier study (Pan et al., 2014b), to determine the zero shear rate viscosities. Values of $\eta_0$ estimated in this manner for the 25, 48.5 and 165.6 kbp DNA in solvent WS are more than an order of magnitude greater, at comparable temperatures and absolute concentrations, than those observed in solvent NS reported earlier in Pan et al. (2018). For instance, at roughly 21.5°C and 0.059 mg/ml, the zero-shear rate viscosity for T4 DNA in solvent WS was 130 mPa.s, while it was 8.9 mPa.s in solvent NS. The values of $\eta_0$ for all the cases measured here are listed in Table 3.

Once the zero shear rate viscosities are determined, it is possible to plot the dependence of the scaled polymer contribution to shear viscosity $\eta_p/\eta_{p0}$ on the raw shear rate $\dot{\gamma}$, for 25 kbp, 48.5 kbp and 165.6 kbp DNA, as shown in various subfigures of Fig. 2. Each subfigure in Fig. 2 belongs to a particular absolute concentration, while the specific symbols indicated in the legends denote different temperatures. The reason the scaled viscosity appears to increase with temperature is because of the division of $\eta_p$ by $\eta_{p0}$. While, as expected, $\eta_p$ decreases with increasing temperature, the shear rate dependence of η_p is more pronounced over the same range of shear rates, for solutions at a lower temperature.

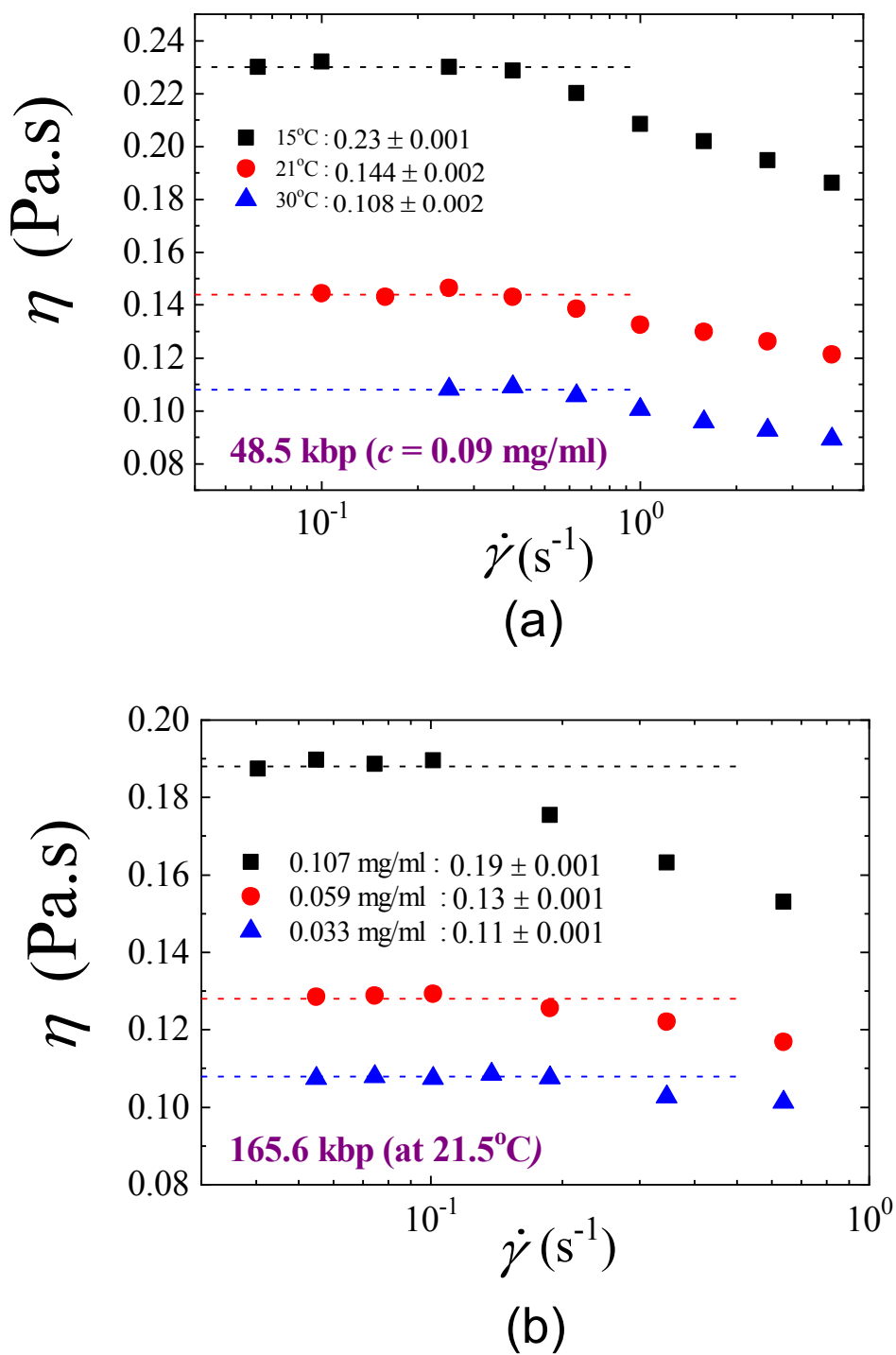

**Fig. 1.** Determination of the zero-shear rate solution viscosity $\eta_0$. The shear rate ($\dot{\gamma}$) dependence of solution viscosity $\eta$ in the region of low shear rate is extrapolated to zero-shear rate (a) for $\lambda$-DNA (48.5 kbp) at a fixed concentration, for a range of temperatures and (b) for T4 DNA (165.6 kbp) at a fixed temperature, for a range of concentrations. The extrapolated values in the limit of zero-shear rate are indicated in the legends.

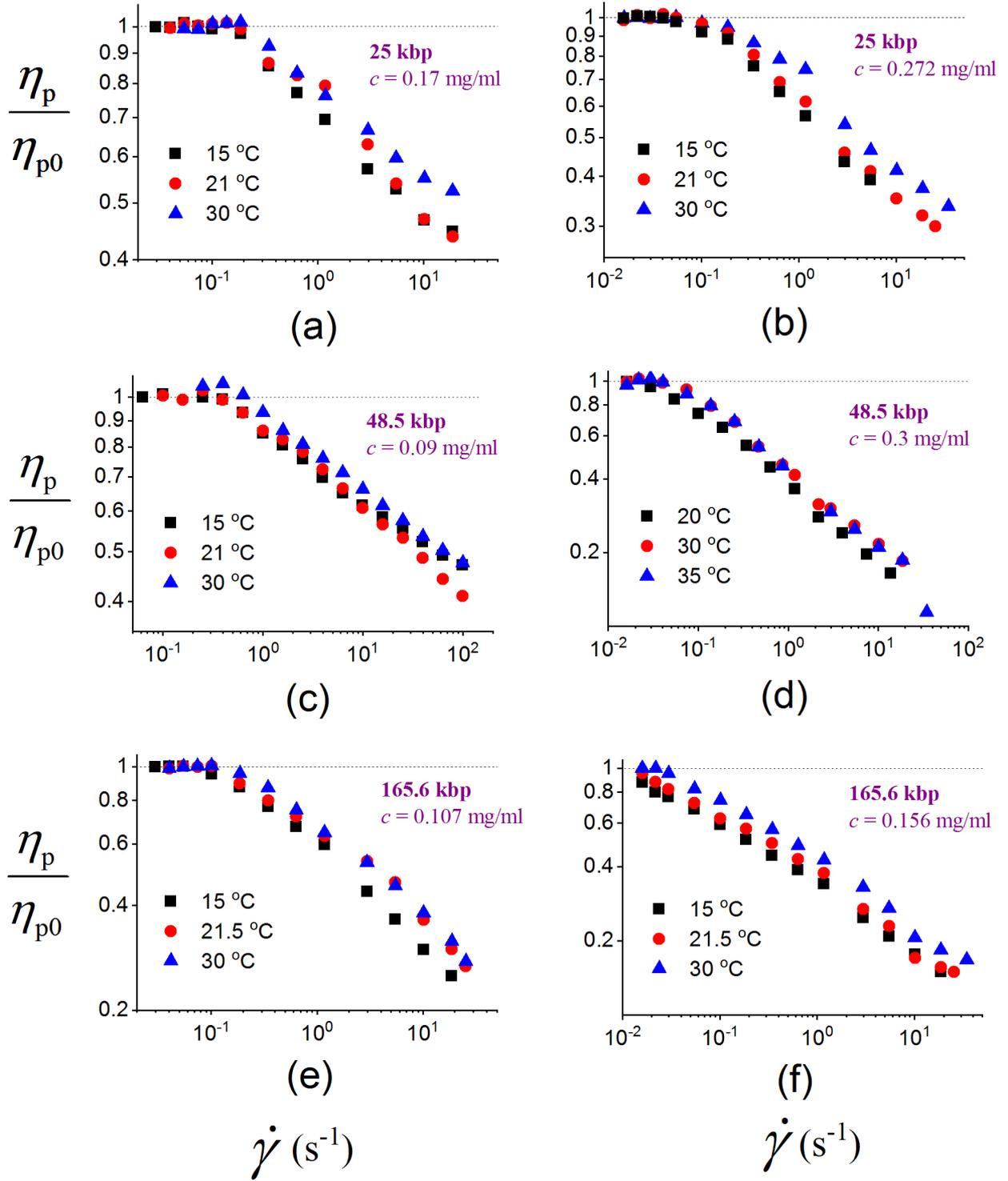

**Fig. 2.** The scaled viscosity $\eta_p / \eta_{p0}$ as a function of the shear rate $\dot{\gamma}$, for semidilute solutions of 25 kbp DNA [subfigures (a) and (b)], $\lambda$-DNA (48.5 kbp) [subfigures (c) and (d)], and T4 DNA (165.6 kbp) [subfigures (e) and (f)], each at a fixed absolute concentration and at different temperatures. The temperatures, concentrations and DNA molecular weight are mentioned in the legends of the individual subfigures. The values of $\eta_0$ corresponding to various temperatures are listed in Table 3.

By defining a concentration dependent large scale relaxation time $\lambda_\eta$,

$$\lambda_\eta = \frac{M\eta_{p0}}{cN_A k_B T} \qquad (1)$$

where $k_B$ is the Boltzmann's constant and $T$ is the absolute temperature, it is possible to collapse data at different temperatures, but at the same concentration, when $\eta_p/\eta_{p0}$ is plotted as a function of $\lambda_\eta \dot\gamma$ as shown in the three subfigures of Fig. 3, corresponding to the three DNA (25, 48.5 and 165.6 kbp) in solvent WS. This implies that using $\lambda_\eta$ as the relaxation time to non-dimensionalize the shear rate leads to time-temperature superposition. In other words, all the different curves seen in the individual subfigures of Figs. 2 collapse on to a single curve for each concentration, independent of the temperature. It is worth noting that since the overlap concentration $c^*$ is a function of temperature, the collapse of data at different temperatures but at the same concentration implies that data on the same curve are at different values of the scaled concentration $c/c^*$. The data collapse is strikingly similar to that reported earlier in Pan et al., 2018 for DNA dissolved in solvent NS, and discussed there in some detail. As in that case, there is a significant shear thinning power law regime over several decades of $\lambda_\eta \dot\gamma$, with the magnitude of the slope increasing with increasing concentration. Interestingly, as pointed earlier in Pan et al. (2018), the slope in the power-law region is not close to –0.5, which was observed previously in the experiments of Hur et al. (2001) and in the simulations of Stoltz et al. (2006) and Huang et al. (2010), where the Weissenberg number was defined in terms of the longest relaxation time $\lambda_1$.

It is instructive to compare the behaviour of the same samples of DNA at 21°C and roughly identical concentrations in the two different solvents WS and NS, as shown in Figs. 4 and 5. It is clear from

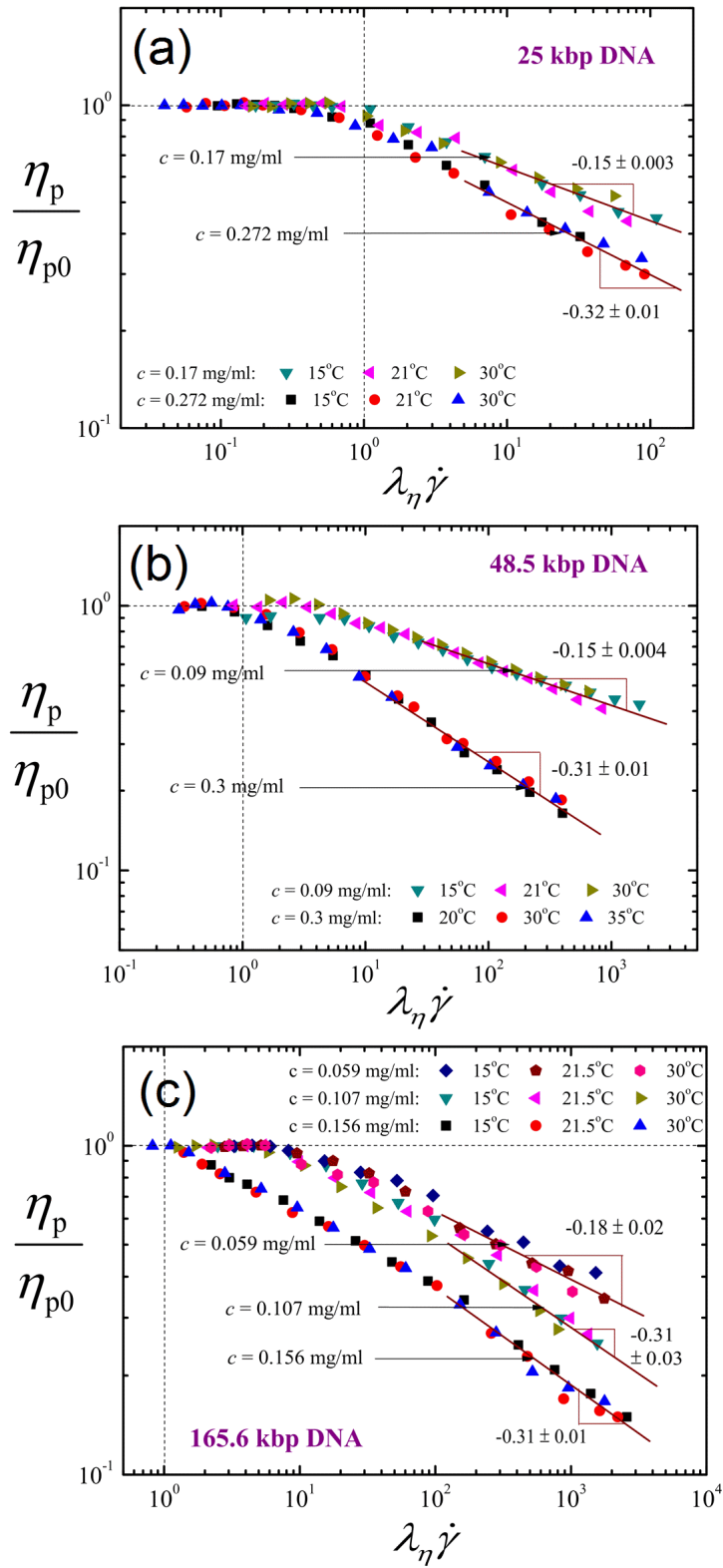

**Fig. 3**. Dependence of the scaled polymer contribution to shear viscosity $\eta_p / \eta_{p0}$, on the Weissenberg number, $\lambda_\eta \dot\gamma$, at different temperatures and concentrations, for linear (a) 25 kbp DNA (b) $\lambda$-DNA (48.5 kbp) and (c) T4 DNA (165.6 kbp) in the solvent WS used for extension studies. The lines are the least squares regression fits to the data in the shear thinning region.

Figs. 4 that even when solutions are at the same temperature, concentration, and non-dimensional shear rate $\lambda_\eta \dot\gamma$, they do not have the same value of $\eta_p/\eta_{p0}$ when the solvent is not the same. DNA solutions in solvent NS shear thin significantly more than solutions in solvent WS at the same value of $\lambda_\eta \dot\gamma$. While the magnitude of the shear thinning exponent increases steadily with increasing concentration in both the solvents, as seen in Fig. 5, the terminal slopes of the viscosity versus $\lambda_\eta \dot\gamma$ curves for samples of DNA in the solvent WS are lower than those for solutions in the solvent NS, at all concentrations. Notably, the asymptotic value of 0.5 is never reached for samples in solvent WS for the concentrations and Weissenberg numbers examined here.

In the light of the significant difference of the shear rheology in these two different solvents, it is worth pointing out that earlier research has shown that the presence of excess sucrose in aqueous DNA solutions can cause a series of structural modifications of DNA chains, including sugar-phosphate and sugar-DNA base bindings (via G-C base pairs), as well as a partial conformational transition from B to A-form, mediated via reorganizations of sucrose intermolecular H-bonding mechanisms (Tajmir-Riahi et al., 1994). Whether sucrose-DNA interactions are responsible for the observed difference in behaviour requires careful further investigation, which was outside the scope of this work.

Pan et al. (2018) have shown previously that by defining a non-dimensional shear rate based on a large-scale relaxation time with a dependence on ($c/c*$) that is different from that of $\lambda_\eta$, namely that of a single correlation blob, it is possible to obtain data collapse of $\eta_p/\eta_{p0}$ at high values of shear rate in the shear thinning regime, across all temperatures and concentrations. Since we do not have measurements of the radius of gyration of DNA in solutions with solvent WS, we are unable to calculate the value of $c^*$ precisely, and carry out a similar analysis here. Additionally, the issue of the influence of sucrose-DNA interactions would have to be borne in mind while carrying out measurements of the radius of gyration. An alternative approach to obtain an estimate of the scaled concentration is discussed in the next section.

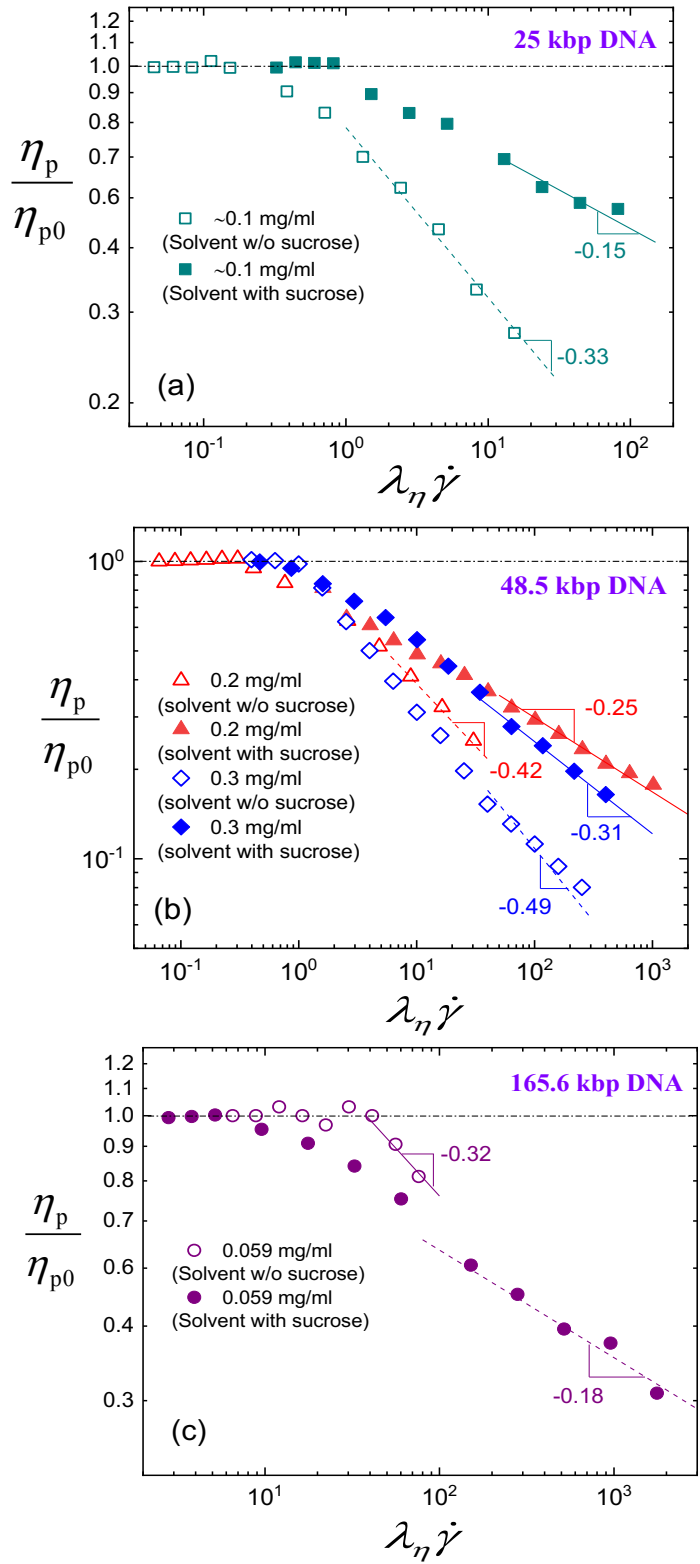

**Fig. 4**. Comparison of shear thinning of DNA chains at nearly same absolute concentration in solvent NS (no sucrose) denoted by open symbols (Pan et al., 2018), and solvent WS (with sucrose, current work), denoted by filled symbols. Representative data have been selected from Fig. 3 at ~21°C for the three different DNA molecular weights.

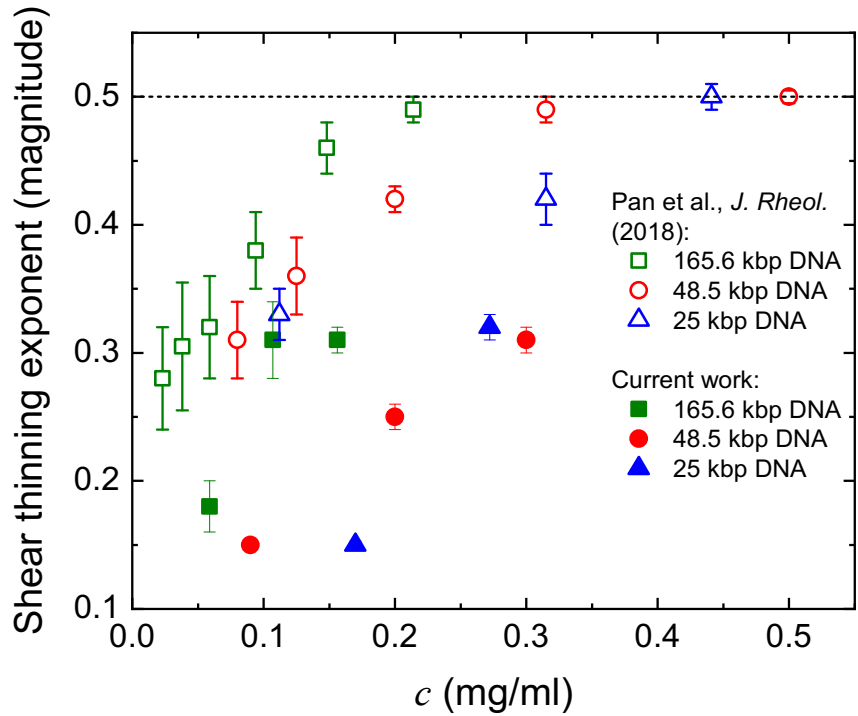

**Fig. 5**. Shear thinning exponent, i.e., the terminal slope of $\eta_p / \eta_{p0}$ versus $\lambda_\eta \dot{\gamma}$, for the three different DNA molecular weights (filled symbols), as a function of absolute concentration $c$, obtained from the curves in Fig. 3. The results are compared with those of the same three DNA (open symbols) from our recent work (Pan et al., 2018). The solvent used in the current work (see Table 2) is the same as in Pan et al. (2018), including identical buffer compositions (10 mM Tris, 1 mM EDTA) and salt (0.5 M NaCl), except with an additional incorporation of ~61 wt.% sucrose.

### 3.2. Uniaxial extensional stress growth in semidilute DNA solutions

Measurements of extensional viscosity are typically carried out using the 'master-plot' technique developed by Gupta et al. (2000), which enables a pre-determined strain-rate ($\dot{\varepsilon}$) to be imposed on the fluid. This method is based on the expectation that for each fluid there is an exclusive relationship between the filament length ($L$) and the mid-filament diameter ($D$) that can be expressed as:

$$\frac{L}{L_0} = f\left(\frac{D}{D_0}\right) \quad (2)$$

where $L_0$ and $D_0$ are the initial values of the length and diameter, respectively. It has been shown that for any desired record of strain-rate and the equivalent diameter-time profile, the length-time profile can be obtained using Eq. (2). The use of these master plots compensates for the non-ideality in extensional flow due to end-effects, and makes data analysis considerably straightforward (Gupta et al., 2000). This procedure has been used extensively in the current work to obtain the diameter profiles for all DNA molecular weights at all the concentrations, and at a wide range of strain rates for each concentration. Figures 6(a)–(d) show the results of using such a technique for the highest concentrations for all the DNA molecular weights, and demonstrates the constancy of the representative values of strain rate $\dot{\varepsilon}$. Figures 6(e)–(h) show the time evolution of the transient Tr$^+$ at different strain rates, where the transient Trouton ratio, Tr$^+$, is defined as the ratio of the extensional stress growth coefficient to the steady zero shear rate viscosity:

$$\text{Tr}^+ = \frac{\eta_E^+(\dot{\varepsilon},t)}{\eta_0} \tag{3}$$

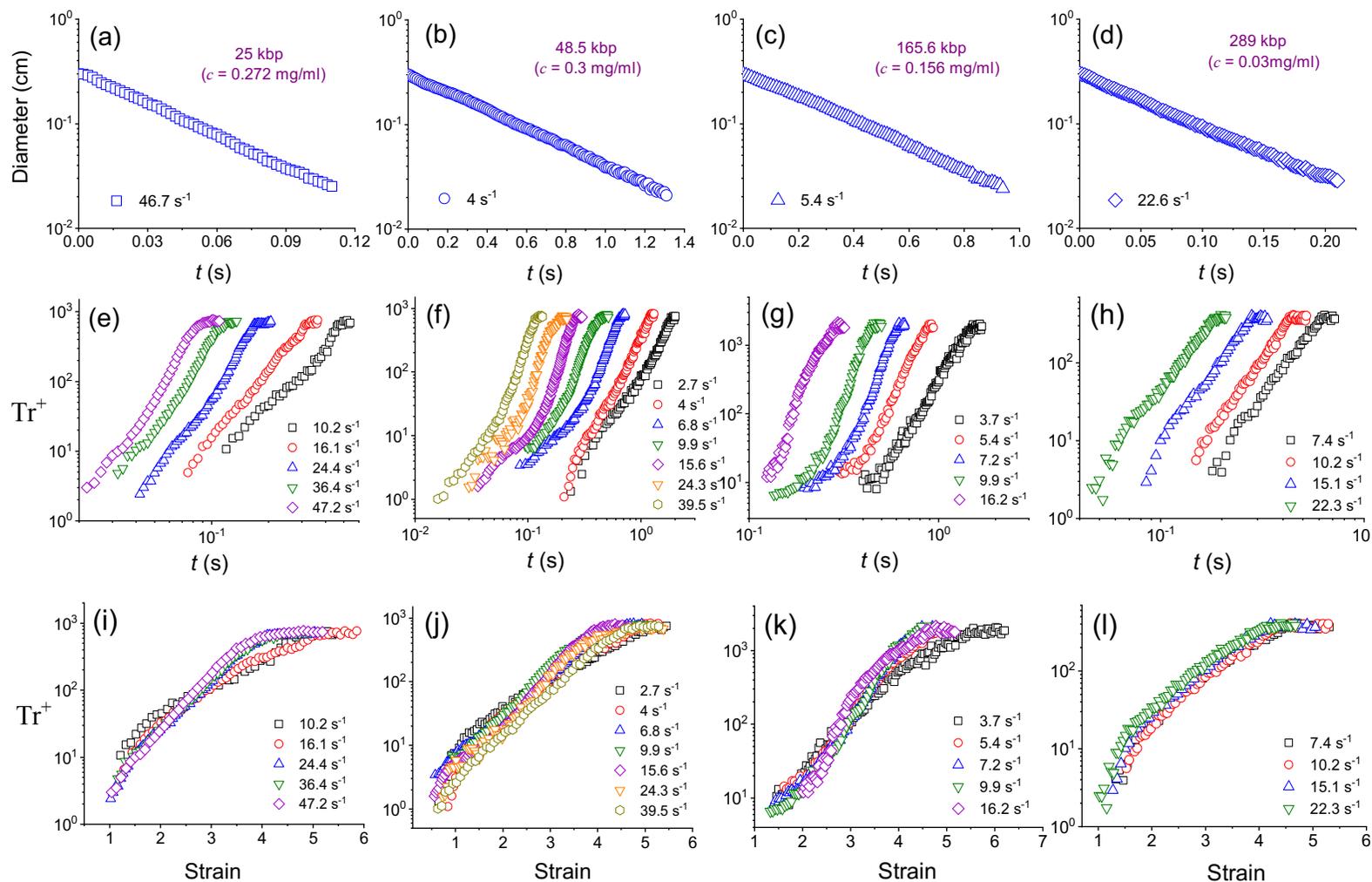

**Fig. 6**. Measurement and extensional characterization profiles for different DNA molecular weights. As indicated in the legends to subfigures (a)–(d), the first, second, third and fourth columns represent the four DNA, in order of increasing molecular weights respectively, and at their highest absolute concentrations. (a)–(d): Diameter profiles obtained using the master-plot technique at different strain rates $\dot{\varepsilon}$ (indicated in the legends); (e)–(h): Dimensionless transient Trouton ratio $\mathrm{Tr}^+$ at different strain rates $\dot{\varepsilon}$ (indicated in the legends); (i)–(l): Asymptotic nature of the experimental normalized elongational stress growth coefficient as a function of strain ($\dot{\varepsilon}\,t$). The strain rates ($\dot{\varepsilon}$) are indicated in the legends.

It is clear that the elongational stress growth can be measured at earlier times and the growth is more rapid, with increasing strain rate, irrespective of chain length. The strain rates used in the current work were chosen such that a uniform elongation of the filament was enforced. These strain rates exceed the inverse viscosity dependent relaxation time $\lambda_\eta$ of the DNA chains used in the sucrose dominated solvent. Lower strain rates than the ones used in the current study were not investigated in order to circumvent the experimental artefacts and instabilities that constrain the procedure (McKinley and Sridhar, 2002). A general observation was that solutions with lower concentrations (not shown in Figs. 6) required high strain rates to overcome the gravitational bending of the filaments, in agreement with the observations of Sunthar et al. (2005) on dilute DNA solutions. Figures 6(i)–(l) show the extensional stress growth data as a function of Hencky strain $\varepsilon$, which is a product of the strain-rate $\dot{\varepsilon}$ and time $t$. The strain-hardening for semidilute DNA solutions seen in these figures is observed for all the DNA molecular weights, at all concentrations (although not shown here), when subjected to uniaxial extensional deformation. The data for different strain rates appear to nearly collapse on top of each other suggesting that the temporal evolution of the extensional stress growth is becoming independent of strain rate, i.e., it is reaching asymptotic values for Wi = $\lambda_\eta \dot{\varepsilon}$ >>1. This is in accordance with previous experiments with dilute DNA and polystyrene solutions (Gupta et al., 2000; Sunthar et al., 2005). Note that the Weissenberg numbers corresponding to the minimum and maximum strain rates used here are listed in Table 3. The Trouton ratio also clearly attains a steady state value at high strains; roughly after 4 strain units for the shorter DNA chains (25 and 48.5 kbp) and after 4.5 strain units for the longer DNA chains (165.6 and 289 kbp). In other words, the different solutions attain steady state after nearly the same number of strain units.

The steady state Trouton ratio Tr in the limit t → ∞ is defined in terms of the steady state extensional viscosity $\eta_E$:

$$\mathrm{Tr} = \frac{\eta_E}{\eta_0} \qquad (4)$$

Measured values of the steady state Trouton ratio as a function of the strain rate, for all the four molecular weight DNA chains considered here, at different concentrations, are shown in the various subfigures of Fig. 7. The constancy of Tr, independent of $\dot{\varepsilon}$, suggests that the extensional stress growth of all the DNA molecules (with an order of magnitude difference in chain lengths), has attained an asymptotic value at the Weissenberg numbers at which the experiments were carried out.

Figures 7 indicate that Tr increases with increasing concentration. This is seen more transparently in Fig. 8, where the dependence of steady state uniaxial extensional viscosities $\eta_E$ on concentration is displayed. Measured values of $\eta_E$, corresponding to different DNA molecular weights, are also listed in Table 3, along with steady state values of Tr and the measured ranges of Wi. To our knowledge, these are the first reported measurements of the steady state uniaxial extensional viscosities, measured with the FSR, of semidilute DNA solutions.

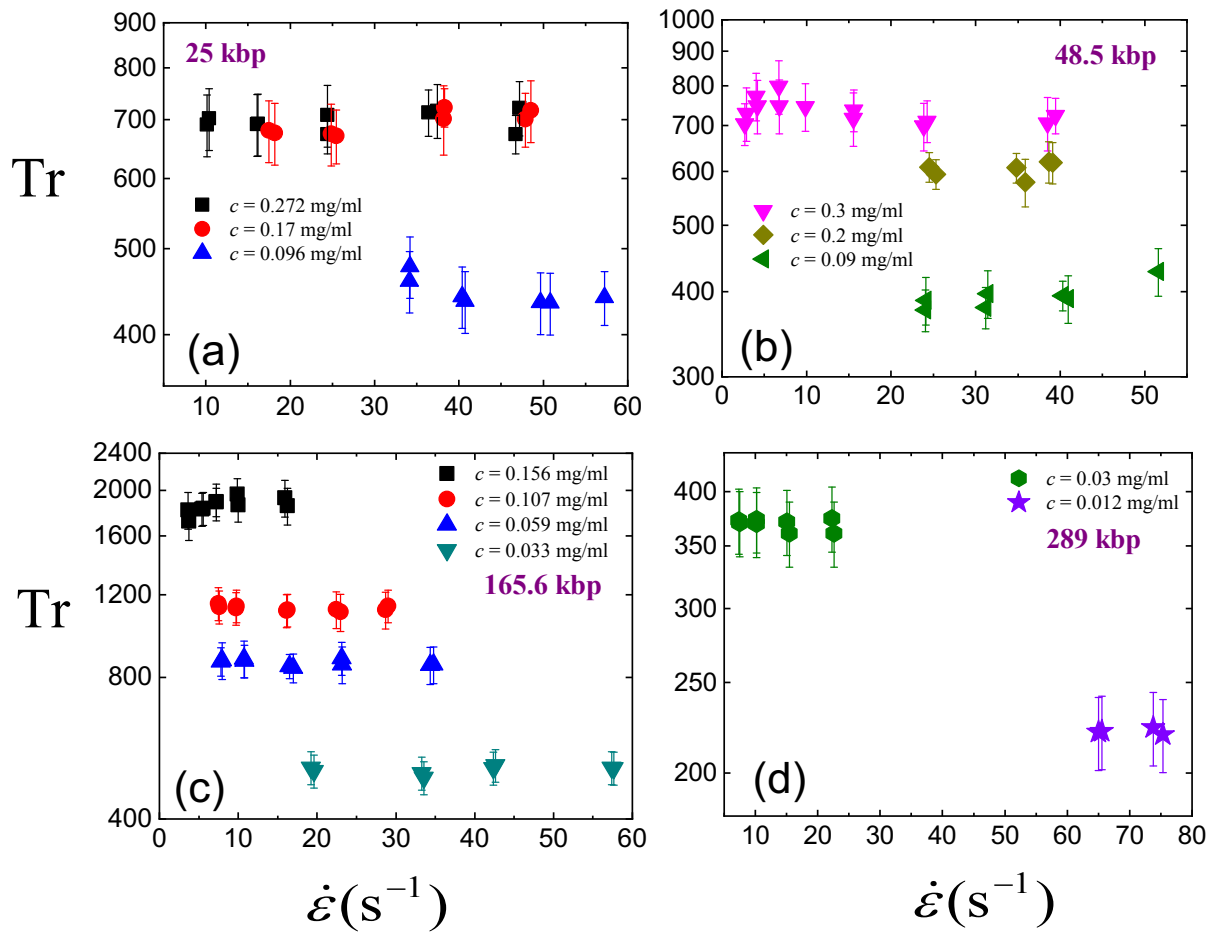

**Fig. 7**. Steady state Trouton ratio, Tr, as a function of strain rate $\dot{\varepsilon}$ for different concentrations (indicated in the legends) of the DNA molecular weights (indicated in the individual subfigures). The percentage errors calculated for each strain rate from the steady state time-averaged Tr are indicated on the data.

It is clear that the extensional viscosity increases quite significantly with increasing concentration. In the case of steady simple shear flow, there are well known scaling laws that describe the dependence of the zero-shear rate viscosity on the scaled concentration *c/c\** in theta and athermal solvents, derived from blob scaling arguments (Rubinstein and Colby, 2003). These have been validated and extended to the double crossover regime by Pan et al. (2014b) for semidilute DNA solutions. Similar scaling arguments do not currently exist for steady state extensional viscosities. It is nevertheless interesting to examine if data collapse can be achieved when the data in Fig. 8 can be reinterpreted in terms of *c/c\**.

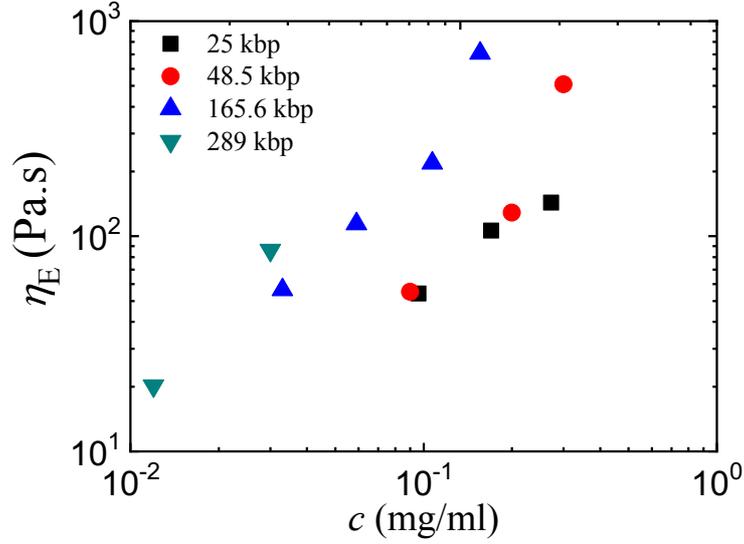

**Fig. 8**. Steady state uniaxial extensional viscosity $\eta_E$ as a function of absolute concentration $c$ for DNA of different molecular weights. A power-law dependence on DNA concentration is evident, along with the effect of DNA molecular weight on $\eta_E$.

As mentioned earlier, values of $c^*$ for DNA solutions in solvent WS are presently unknown since the radius of gyration as a function of temperature and molecular weight is not known. On the other hand, even though the theta temperature in this solvent is not known, one can determine the overlap concentration at the theta temperature $c_\theta^*$, since the radius of gyration at the theta temperature $R_g^\theta$, is known for any DNA from a knowledge of its contour length and the number of Kuhn steps in the chain. Values of $R_g^\theta$ for all the DNA molecular weights $M$ used in the present work are listed in Table 1. It follows that one can calculate the overlap concentration $c_\theta^*$ from the expression (Rubinstein and Colby, 2003):

$$c^* = \frac{3M}{4\pi N_A R_g^3} \tag{5}$$

where $N_A$ is the Avogadro number. Values of $c_\theta^*$ determined in this manner are given in Table 3 for all the DNA. Figure 9 (a) is a reinterpretation of the data in Fig. 8, where the dependence of the scaled polymer contribution to steady state extensional viscosity, $(\eta_E - 3\eta_s)/\eta_s$, is displayed as a function of the scaled concentration $c/c_\theta^*$. Clearly there is much less scatter in the data when compared to Fig. 8, with values for different chain lengths lying much closer to each other, with an apparent power law dependence of the scaled extensional viscosity on the scaled concentration. This trend towards data collapse when interpreted in terms of a scaled concentration, encouraged us to try to estimate the value of $c^*$ that leads to the best collapse of the steady state extensional viscosity data, using the following procedure.

**Table 3**: Steady state uniaxial extensional viscosities ($\eta_E$), zero shear rate viscosities ($\eta_0$), Trouton ratios (Tr) and Weissenberg numbers (Wi) for all the DNA samples used in this work at different absolute concentrations $c$. All measurements were carried out at 21±0.5 °C. An indicative range of Wi (between $\text{Wi}_{\min}$ and $\text{Wi}_{\max}$) is shown for various strain rates $\dot{\varepsilon}$ used. Here, $\text{Wi} = \lambda_\eta \, \dot{\varepsilon}$, where $\lambda_\eta$ have been estimated from $\eta_{p0}$ (= $\eta_0 - \eta_s$) values in this solvent using Eq. (1). Trouton ratio Tr values have been calculated using Eq. (4). The solvent viscosity $\eta_s$ at 21°C was used from Table 2. The values of overlap concentration $c_\theta^*$ have been estimated based on the analytical $R_g^\theta$ values from Table 1 using Eq. (5). The values of $z$ and ($c/c^*$) provide the best collapse of the steady state extensional viscosity data, as displayed in Fig. 9 (c).

| DNA Size (kbp) | $c$ (mg/ml) | $c_\theta^*$ (mg/ml) | $c/c_\theta^*$ | $z$ | $c/c^*$ | $\eta_0$ (Pa.s) | $\eta_E$ (Pa.s) | $\lambda_\eta$ (s) | Tr | $\text{Wi}_{\min}$ | $\text{Wi}_{\max}$ |
|---|---|---|---|---|---|---|---|---|---|---|---|
| 25 | 0.096 | 0.123 | 0.78 | 0.5 | 1.29 | 0.122 | 54.2 | 4.3 | 446.4 | 146 | 245 |
|  | 0.17 |  | 1.38 |  | 2.28 | 0.154 | 106.3 | 3.7 | 692.6 | 65 | 179 |
|  | 0.272 |  | 2.21 |  | 3.65 | 0.204 | 143.1 | 3.6 | 698.2 | 36 | 168 |
| 48.5 | 0.09 | 0.089 | 1.01 | 0.7 | 1.89 | 0.144 | 55.2 | 6.3 | 393.3 | 151 | 323 |
|  | 0.2 |  | 2.25 |  | 4.21 | 0.213 | 128.8 | 5.2 | 604.2 | 127 | 202 |
|  | 0.3 |  | 3.37 |  | 6.32 | 0.693 | 508.9 | 14.3 | 734.1 | 39 | 564 |
| 165.6 | 0.033 | 0.048 | 0.69 | 1.3 | 1.65 | 0.110 | 56.4 | 10.1 | 512.7 | 194 | 579 |
|  | 0.059 |  | 1.23 |  | 2.95 | 0.130 | 114.2 | 8.0 | 878.5 | 62 | 273 |
|  | 0.107 |  | 2.23 |  | 5.34 | 0.190 | 218.7 | 8.2 | 1151.1 | 61 | 235 |
|  | 0.156 |  | 3.25 |  | 7.79 | 0.382 | 709.2 | 13.9 | 1856.5 | 52 | 226 |
| 289 | 0.012 | 0.036 | 0.33 | 1.7 | 0.91 | 0.091 | 20.2 | 16.9 | 221.7 | 1104 | 1279 |
|  | 0.03 |  | 0.83 |  | 2.26 | 0.226 | 86.2 | 37.4 | 369.3 | 277 | 1922 |

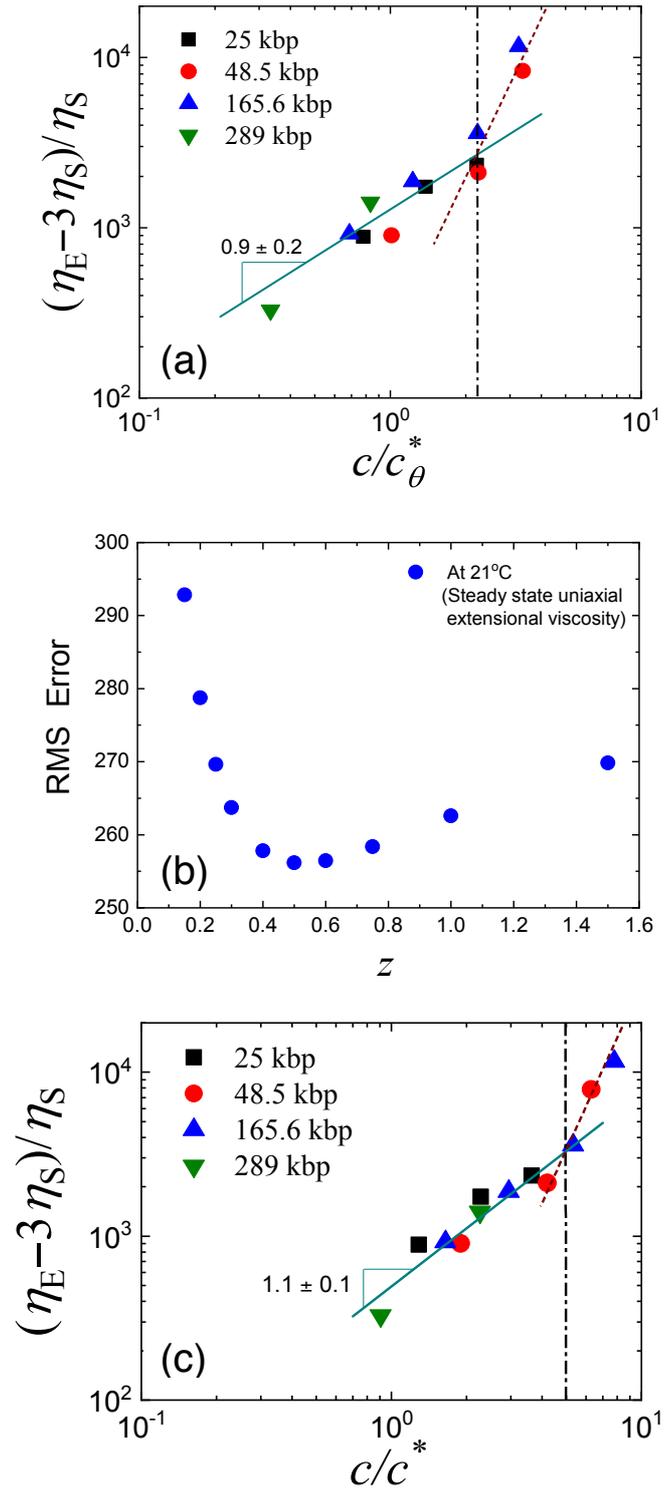

**Fig. 9**. Dimensionless polymer contribution to steady state uniaxial extensional viscosity $\eta_E$ as a function of normalized concentration for DNA of different molecular weights: (a) data collapse with $c/c_\theta^*$ as the independent variable, (b) RMS error (see Eq. (8)) in the power law fit to extensional viscosity data as a function of solvent quality $z$, and (c) data collapse with $c/c^*$ as the independent variable, at the lowest value of the RMS error. The solid lines in both subfigures represent least-squares regression fits to the data.

In order to find the radius of gyration $R_g$ of DNA at any temperature and molecular weight, one needs to know the solvent quality $z$, defined by

$$z = k \left(1 - \frac{T_\theta}{T}\right) \sqrt{M} \qquad (6)$$

where, $k$ is a chemistry dependent constant and $T_\theta$ is the theta temperature. Once $z$ is known, the swelling of the radius of gyration $\alpha_g = (R_g/R_g^\theta)$ can be determined from the universal swelling function $\alpha_g = \alpha_g(z)$, which has the following form (Pan et al. 2014a),

$$\alpha_g = (1 + az + bz^2 + cz^3)^{m/2} \qquad (7)$$

with, $a = 9.5286, b = 19.48 \pm 1.28, c = 14.92 \pm 0.93, m = 0.133913 \pm 0.0006$. The values of $k$ and $T_\theta$ for DNA in the solvent NS have been determined by Pan et al. (2014a, 2014b), who have also shown that DNA in the presence of excess salt exhibits the same universal scaling behaviour as neutral synthetic polymer solutions. In the case of DNA in solvent WS, even though we don't know both $k$ and $T_\theta$, we can find $R_g$ by guessing a value of $z$, and using the function $\alpha_g = \alpha_g(z)$ given in Eq. (7). Note that we only need to guess $z$ for a single DNA molecular weight, say $M_1$, since at any other molecular weight $M_2$, the solvent quality $z_2 = z_1(\sqrt{M_2/M_1})$, when considering both molecular weights at the same temperature.

By guessing a value of $z$ for 25 kbp DNA, we have calculated $\alpha_g$ from Eq. (7), then determined $R_g$ since $R_g^\theta$ is known, and finally estimated $c^*$ at the guessed value of $z$ from Eq. (5). Knowing $z$ for 25 kbp DNA, the values of $z$ for all the other DNA molecular weights can be determined as discussed above, and finally the values of $c^*$ at the respective values of $z$ can be estimated as was done for 25 kbp DNA. Measured values of the scaled viscosity $(\eta_E - 3\eta_s)/\eta_s$ can then be plotted as a function of $(c/c^*)$ for all the DNA samples. By fitting a power law to the experimental data (leaving out the last two data points at high concentrations for $\lambda$-phage and T4 DNA that appear to be in the entangled regime), the root mean square error between the fit and experimental data was estimated from,

$$\text{RMS Error} = \sqrt{\frac{1}{N}\sum_{i=1}^{N}\left(Y_i^{Pred} - Y_i^{Observed}\right)^2} \qquad (8)$$

where $Y = (\eta_E - 3\eta_s)/\eta_s$. Figure 9 (b) is a plot of RMS Error as a function of $z$ for 25 kbp DNA. It is clear that the least error occurs for $z = 0.5$. Using this value of $z$ for 25 kbp DNA, the estimated values of $z$ for all the other DNA are listed in Table 3, along with the corresponding values of $c/c^*$. Figure 9 (c) is a plot of $(\eta_E - 3\eta_s)/\eta_s$ versus $(c/c^*)$ for these values of $z$. Clearly, the data collapse is quite compelling, and certainly better than that displayed in Fig. 9 (a) when using $c_\theta^*$ as the scaling variable for concentration. Interestingly, the estimated value of $z = 0.7$ at 21°C for $\lambda$-phage DNA is very close to the value used by Sasmal et al., (2017) in their Brownian dynamics simulations, which were aimed at predicting the planar extensional flow measurements by Hsiao et al., (2017a) of semidilute DNA solutions in a cross-slot device. It remains to be seen if the estimated values of $z$ and $\alpha_g$ that provide the best fit to the steady state extensional viscosity data turn out to be accurate when the actual values of $k$, $T_\theta$ and $R_g$ are determined for DNA in solvent WS by direct experimental measurements. It would also greatly improve our understanding if the collapse could be explained within the framework of a scaling theory.

It is worth noting from Fig. 9 (c) that there appears to be a change in slope between $c/c^* \approx 4$ and 5, suggesting a change in the scaling regime from a semidilute unentangled to the entangled regime. A change in slope was observed previously by Pan et al. (2014b) at roughly similar values of $c/c^*$ in the scaling of the zero-shear rate viscosity with the scaled concentration.

## 4. Conclusions

Steady shear viscosities have been measured for semidilute unentangled DNA solutions of three different molecular weights (25, 48.5 and 165.6 kbp), for a wide range of concentrations and temperatures, in a solvent with a high concentration of sucrose. All the DNA molecular weights, at all the concentrations investigated in the current study, demonstrated a predominantly shear thinning behaviour at high shear rates. Time-temperature superposition was confirmed when the data was interpreted in terms of $\lambda_\eta \dot{\gamma}$. However, data at different absolute concentrations could not be collapsed, as was observed recently by Pan et al. (2018), in a solvent without sucrose. The shear thinning exponents in the solvent with sucrose were of lower magnitude, at similar concentrations, when compared to values in the solvent without sucrose. The possibility that high concentrations of sucrose might lead to structural modifications of DNA and that sucrose-DNA interactions might influence the shear-thinning behaviour of semidilute DNA solutions is worthy of further investigation.

Elongational stress growth has been measured using a filament stretching rheometer, for semidilute unentangled DNA solutions of four different molecular weights (25, 48.5, 165.6 and 289 kbp), by subjecting them to uniaxial extensional flows for a wide range of concentrations, at room temperature, in a solvent which is predominantly sucrose. The transient Trouton ratio increased steadily with increasing strain rate at similar values of time after inception of flow. For each DNA molecular weight, existence of a steady state was confirmed between 4–4.5 strain units, irrespective of concentration. All DNA molecular weights approached asymptotic values of elongational stress growth expected at high Weissenberg numbers. The scaled polymer contribution to the steady state extensional viscosity shows a rough collapse on a master curve when interpreted in terms of the scaled concentration $c/c^*$, with an apparently power law dependence. It is envisaged that the current experimental observations will serve as benchmark data for the extensional viscosity of semidilute unentangled DNA as a function of molecular weight and concentration, and prove to be useful for validations of theoretical predictions and numerical simulations.

## Acknowledgement

This work was supported by the ARC Discovery Projects grant arrangement (Project No. DP120101322). Douglas Smith at UCSD and Brad Olsen at MIT are thanked for their generous gifts of the originally synthesized 25 kbp Fosmid and 289 kbp BAC DNA constructs as agar stab cultures of *E. coli*. Authors would like to acknowledge Michael Danquah (previously at Monash University), for facilitating lab space as well as DNA isolation amenities. IIT-B Monash research Academy at IIT Bombay, India is thanked for funding and overall provisions. We also thank the anonymous referee for helpful suggestions that have improved the quality of the paper.